\documentclass[a4paper,draftcls,11pt,onecolumn]{IEEEtran}

\usepackage{amsmath}
\usepackage{amssymb}
\usepackage{amsthm}

\DeclareMathOperator{\E}{E}

\newcommand{\es}{\varepsilon_s}
\newcommand{\capPois}{\rm C}

\newtheorem{theorem}{Theorem}

\title{Achievability of the Rate $\frac{1}{2}\log(1+\es)$ in the Discrete-Time Poisson Channel}
\author{Alfonso Martinez%
  \thanks{A.\ Martinez is with Centrum Wiskunde \& Informatica,  The Netherlands. 
  e-mail: alfonso.martinez@ieee.org.}}
\begin{document}

\maketitle

\begin{abstract}
A simple lower bound to the capacity of the discrete-time Poisson channel with average energy $\es$ is derived. The rate $\frac{1}{2}\log(1+\es)$ is shown to be the generalized mutual information of a modified minimum-distance decoder, when the input follows a gamma distribution of parameter $1/2$ and mean $\es$. 
\end{abstract}

\section{Introduction}
Consider a memoryless discrete-time whose output $Y$ is distributed according to a Poisson distribution of parameter $X$, the channel input. By construction, the output is a non-negative integer, and the input a non-negative real number. The channel transition probability $W(y|x)$ is thus given by
\begin{equation}
	W(y|x) = e^{-x}\frac{x^y}{y!}.
\end{equation}
This model, the discrete-time Poisson (DTP) channel, appears often in the analysis of optical communication channels. In this case, one can identify the input with a signal energy and the output with an integer number of  quanta of energy.   

Let $P_X(x)$ denote the probability density function of the channel input. 
We assume that the input energy is constrained, i.\ e.\ $\E[X]\leq \es$, where $\E[\cdot]$ denotes the expectation operator and $\es$ is the average energy. Random variables are denoted by capital letters, and their realizations by small letters. 
  
An exact formula for the capacity $\capPois(\es)$ of the DTP channel is not known. 
%
Recently, Lapidoth and Moser \cite{lapidoth03:boundsCapacityPoisson},  derived the following lower bound
\begin{align}\label{eq:lowPoissonlm}
    \capPois(\es) &\geq \log\left(\biggl(1+\frac{1}{\es}\biggr)^{ 1+\es}\sqrt{\es}\right) -\biggl(1+\sqrt{\frac{\pi}{24\es}}\biggr). 
\end{align}
Observe that this bound diverges for vanishing $\es$.
Capacity is given in nats and the logarithms are in base $e$.

A closed-form expression for the mutual information $I(X;Y)$ achieved by an input with a gamma distribution of parameter $\nu$ was derived by Martinez in \cite{martinez07:spectralEfficiencyOpticalDirectDetection}, namely 
\begin{align}\label{eq:mInfGamma}
I(X;Y) &=  \int_0^1 \Biggl(\es-\biggl(1-\frac{\nu^\nu}{(\nu+\es (1-u))^\nu}\biggr)\frac{u^{\nu-1}}{1-u}\Biggr)\frac{du}{\log u}\notag \\&\qquad +(\es+\nu)\log\frac{\es+\nu}{\nu}  + \es \bigl(\psi(\nu+1)-1\bigr),
\end{align}
where $\psi(y)$ is Euler's digamma function. 
For $\nu = 1/2$, numerical evaluation of the mutual information gives a rate which would seem to exceed $\frac{1}{2}\log(1+\es)$ for all values of $\es$. In this paper, we prove that the rate $\frac{1}{2}\log(1+\es)$ is indeed achievable by this input distribution. The analysis uses a suboptimum minimum-distance decoder, similar in spirit to Lapidoth's analysis of nearest neighbor decoding \cite{lapidoth96:nearestNeighborDecodingAdditiveNon-GaussianNoiseChannels}. 

\section{Main Result}

Let the input $X$ follow a gamma distribution of parameter $1/2$ and mean $\es$, that is,
\begin{equation}\label{eq:gammadensity}
	P_X(x) = \frac{1}{\sqrt{2\pi \es x}}e^{-\frac{x}{2\es}}.
\end{equation}
This choice led to good lower and upper bounds in \cite{lapidoth03:boundsCapacityPoisson} and \cite{martinez07:spectralEfficiencyOpticalDirectDetection} respectively.

We consider a maximum-metric decoder; the codeword metric is given by the product of symbol metrics $q(x,y)$ over all channel uses.  The optimum maximum-likelihood decoder, for which $q(x,y) = W(y|x)$, is somewhat unwieldy to analyze (Eq.~\eqref{eq:mInfGamma} gives the exact mutual information). We consider instead a symbol decoding metric of the form
\begin{equation}\label{eq:metric}
	q(x,y) = e^{-a x -\frac{y^2}{x}},
\end{equation}
where $a = 1+\frac{1}{\es}$. The reasons for this choice of $a$ will be apparent later. 

Clearly, the decoder is unchanged if we replace the symbol metric $q(x,y)$ by a symbol distance $d(x,y) = -\log q(x,y)$, and select the codeword with smallest total distance, summed over all channel uses. This alternative formulation is reminiscent of minimum-distance, or nearest-neighbor decoding. Indeed, the metric in Eq.~\eqref{eq:metric} is equivalent to a minimum-distance decoder which uses the distance 
\begin{equation}\label{eq:distance}
	d(x,y) = \frac{(y-\sqrt{a}x)^2}{x} = \frac{y^2}{x}+ax-2y\sqrt{a}.
\end{equation}
The term $-2y\sqrt{a}$ is common to all symbols $x$ and can be removed, since it does not affect the decision.

For $a = 1$, the distance in Eq.~\eqref{eq:distance} naturally arises from a Gaussian approximation to the channel output, whereby the channel output is modeled as a Gaussian random variable of mean $x$ and variance $x$. This approximation is suggested by the fact that a Poisson random variable of mean $x$ approaches a Gaussian random variable of mean and variance $x$ for large $x$. 

Minimum-distance decoders were considered by Lapidoth \cite{lapidoth96:nearestNeighborDecodingAdditiveNon-GaussianNoiseChannels} in his analysis of additive non-Gaussian-noise channels. For our channel model, even though noise is neither additive (it is signal-dependent), nor Gaussian, similar techniques to the ones used in \cite{lapidoth96:nearestNeighborDecodingAdditiveNon-GaussianNoiseChannels} can be applied.
More specifically, since we have a {\em mismatched decoder}, we determine the generalized mutual information \cite{ganti00:mismatchedDecodingRevisited}. For a given decoding metric $q(x,y)$ and a positive number $s$, it can be proved \cite{ganti00:mismatchedDecodingRevisited} that the following rate ---the generalized mutual information--- is achievable
\begin{equation}\label{eq:deftGMI}
I_\text{GMI}(s) = \E\left[\log\frac{q(X,Y)^s}{\int P_X(x) q(x',Y)^s\,dx'}\right].
\end{equation}
The expectation is carried out according to $P_X(x)W(y|x)$. This quantity is obviously a lower bound to the channel capacity.

Our main result is
\begin{theorem}\label{thm:1}
In the discrete-time Poisson channel with average signal energy $\es$, the rate $\frac{1}{2}\log(1+\es)$ is achievable.
\end{theorem}
This rate is reminiscent of the capacity of a real-value Gaussian channel with average signal-to-noise ratio $\es$. Similarly to the situation in this channel, the rate is achieved by a form of minimum-distance decoding. Differently, the input follows a gamma distribution, rather than a Gaussian.

\begin{proof}
We evaluate the generalized mutual information $I_\text{GMI}(s)$ for an input distributed according to the gamma density, in Eq.~\eqref{eq:gammadensity}.
First, we evaluate the expectation in the denominator \cite[Eq.~3.471-15]{gradshteyn00:tableIntegrals}
\begin{align}
	\int_0^\infty\frac{e^{-\frac{x'}{2\es}-asx' -\frac{sy^2}{x'}}}{\sqrt{2\pi\es x'}}\,dx' 
	&= \frac{e^{-y\sqrt{\frac{2s(1+2a\es s)}{\es}}}}{\sqrt{1+2a\es s}}.
\end{align}

Further, using the expression of the first two moments of the Poisson distribution, namely\footnote{The moment generating function of a Poisson random variable of mean $x$ is readily computed to be $e^{x(e^t-1)}$. The first two moments are the first two derivatives, evaluated at $t = 0$.}
\begin{align}
	&\sum_y W(y|x) y = x, \quad \sum_yW(y|x) y^2 = x^2 +x,
\end{align}
together with the input constraint $\int P_X(x)x\,dx = \es$, we can explicitly carry out the expectation in Eq.~\eqref{eq:deftGMI},
\begin{align}
	I_\text{GMI}(s) 
	&= \int P_X(x)\sum_y W(y|x) \log\bigl(q(x,y)^s\bigr)\,d x \notag\\
	&\qquad - \int P_X(x)\sum_y W(y|x) \log\left(\int P_X(x')q(x',y)^s\, dx'\right)\,d x\\
	&= s\int P_X(x)\sum_y W(y|x)\left(-a x -\frac{y^2}{x}\right)\,dx \notag\\
	&\qquad - \int P_X(x)\sum_y W(y|x) \left(-y\sqrt{\frac{2s(1+2a\es s)}{\es}}-\log\sqrt{1+2a\es s}
	\right)\,d x \label{eq:14} \\
	&= -s\left((a+1)\es+1\right) + \sqrt{2\es s(1+2a\es s)}+\frac{1}{2}\log(1+2a\es s).
\end{align}

Choosing $\hat{s} = \frac{2\es}{(a-1)^2\es^2+2\es(a+1)+1}$, the first two summands cancel out. And for $a = 1+\frac{1}{\es}$ we have that $2a\hat{s} = 1$, and therefore
\begin{align}\label{eq:iGMIfinal}
	I_\text{GMI}(\hat{s}) &= \frac{1}{2}\log(1+\es).
\end{align}
\end{proof}

The same rate, $\frac{1}{2}\log(1+\es)$, is also achievable by a decoder with $a = 1$. In this case, we have to replace the generalized mutual information by the alternative expression $I_\text{LM}$ \cite{ganti00:mismatchedDecodingRevisited}, given by
\begin{equation}
  I_\text{LM} = \E\left[\log\frac{a(X)q(X,Y)^s}{\int P_X(x) a(x')q(x',Y)^s\,dx'}\right].
\end{equation}
As for $I_\text{GMI}$, $s$ is a non-negative number; $a(x)$ is a weighting function. Setting $a(x) = e^{-\frac{s}{\es}x}$ we have that $I_\text{LM}$ is given by Eq.~\eqref{eq:14}, thus proving the achievability.

The bound provided in this paper is simpler and tighter than Eq.~\eqref{eq:lowPoissonlm}. It would be interesting to extend Theorem~\ref{thm:1} to channel models $Y = S(X) + Z$, where $S(X)$ corresponds to the case considered here and $Z$ is some additive noise $Z$, with a Poisson or a geometric distribution. A different input distribution and another modified decoding metric are likely required for either case.



\end{document}